\documentclass[useAMS, usenatbib]{mn2e}
\input psfig.sty

\newcommand {\apj} {ApJ}
\newcommand {\apjl} {ApJL}
\newcommand {\apjs} {ApJS}
\newcommand {\mnras} {MNRAS}

\newcommand {\aap} {A\&A}
\newcommand {\aj} {AJ}

\newcommand {\nat} {Nature}
\newcommand {\araa} {ARA\&A}

\newcommand {\etal} {et~al.~}
\def \spose#1{\hbox  to 0pt{#1\hss}}  
\newcommand {\lta} {\mathrel{\spose{\lower 3pt\hbox{$\sim$}}\raise  2.0pt\hbox{$<$}}}
\newcommand {\gta} {\mathrel{\spose{\lower  3pt\hbox{$\sim$}}\raise 2.0pt\hbox{$>$}}}
\def \ion#1#2{#1{\footnotesize{#2}}\relax}
\newcommand {\ha}  {\ifmmode H\alpha \else H$\alpha $ \fi} 
\newcommand {\hi} {\ion{H}{I} } 

\newcommand {\kms} {\ifmmode  \,\rm km\,s^{-1} \else $\,\rm km\,s^{-1}  $ \fi }
\newcommand {\kpc} {\ifmmode  {\rm kpc}  \else ${\rm  kpc}$ \fi  }  
\newcommand {\Msun} {\ifmmode M_{\odot} \else $M_{\odot}$ \fi}

\newcommand {\LCDM} {\ifmmode \Lambda{\rm CDM} \else $\Lambda{\rm CDM}$ \fi}
\newcommand {\fbar} {\ifmmode f_{\rm bar} \else $f_{\rm bar}$ \fi}

\newcommand {\Rvir} {\ifmmode R_{\rm vir} \else $R_{\rm vir}$ \fi}
\newcommand {\Vvir} {\ifmmode V_{\rm  vir} \else  $V_{\rm vir}$  \fi} 
\newcommand {\Mvir} {\ifmmode M_{\rm  vir} \else $M_{\rm  vir}$ \fi}  

\newcommand {\Mstar}    {\ifmmode M_{\rm star}    \else $M_{\rm star}$ \fi} 
\newcommand {\Mgas}    {\ifmmode M_{\rm gas}    \else $M_{\rm gas}$ \fi} 
\newcommand {\Mgal}  {\ifmmode M_{\rm gal}  \else $M_{\rm gal}$ \fi}
\newcommand {\Mbar}  {\ifmmode M_{\rm bar}  \else $M_{\rm bar}$ \fi}
\newcommand {\Vflat} {\ifmmode  V_{\rm flat} \else $V_{\rm flat}$ \fi}

\title[BTF and Galactic Outflows]
{The baryonic Tully-Fisher relation and galactic outflows}
    
\author[Dutton]
{Aaron A. Dutton$^{1}$\thanks{dutton@mpia.de}\\
  $^1$Max Planck Institute for Astronomy, K\"onigstuhl 17, 69117, Heidelberg, Germany\\
}

\begin{document}
             
\date{accepted to MNRAS}
             
\pagerange{\pageref{firstpage}--\pageref{lastpage}}\pubyear{2012}

\maketitle           

\label{firstpage}

%%%%%%%%%%%%%%%%%%%%%%%%%%%%%%%%%%%%%%%%%%%%%%%%%%%%%%%%%%%%%%%%%%%%%%

\begin{abstract}
  Most of the baryons in the Universe are not in the form of stars and
  cold gas in galaxies. Galactic outflows driven by supernovae/stellar
  winds are the leading mechanism for explaining this fact.  The
  scaling relation between galaxy mass and outer rotation velocity
  (also known as the baryonic Tully-Fisher relation, BTF) has recently
  been used as evidence against this viewpoint.  We use a \LCDM based
  semi-analytic disk galaxy formation model to investigate these
  claims.
  In our model, galaxies with less efficient star formation and higher
  gas fractions are more efficient at ejecting gas from galaxies.
  This somewhat counter intuitive result is due to the (observational)
  fact that galaxies with less efficient star formation and higher gas
  fractions tend to live in dark matter haloes with lower circular
  velocities, from which less energy is required to escape the
  potential well.
  In our model the intrinsic scatter in the BTF is $\simeq 0.15$ dex,
  and mostly reflects scatter in dark halo concentration. The scatter
  is largely independent of galaxy structure because of the large
  radius within which galaxy rotation velocities are measured.  The
  observed scatter, equal to $\simeq 0.24$ dex, is dominated by
  measurement errors. The best estimate for the intrinsic scatter is
  that it is less than 0.15 dex, and thus our \LCDM based model (which
  does not include all possible sources of scatter) is only just
  consistent with this.  Future observations of the BTF scatter could
  be made with a more stringent measurement of the intrinsic scatter,
  and thus provide a strong constraint to galaxy formation models.
  In our model, gas rich galaxies, at fixed virial velocity ($\Vvir$),
  with lower stellar masses have lower baryonic masses.  This is
  consistent with the expectation that galaxies with lower stellar
  masses have had less energy available to drive an outflow.  However,
  when the outer rotation velocity ($\Vflat$) is used the correlation
  has the opposite sign, with a slope in agreement with
  observations. This is due to the fact there is scatter in the
  relation between $\Vflat$ and $\Vvir$.
  In summary, contrary to some previous claims, we show that basic
  features of the BTF are consistent with a \LCDM based model in which
  the low efficiency of galaxy formation is determined by galactic
  outflows.

\end{abstract}

\begin{keywords}
  galaxies: formation -- galaxies: fundamental parameters -- galaxies:
  haloes -- galaxies: spiral -- galaxies: kinematics and dynamics
\end{keywords}

\setcounter{footnote}{1}

%%%%%%%%%%%%%%%%%%%%%%%%%%%%%%%%%%%%%%%%%%%%%%%%%%%%%%%%%%%%%%%%%%%%%%
%% SECTION 1: INTRODUCTION
%%%%%%%%%%%%%%%%%%%%%%%%%%%%%%%%%%%%%%%%%%%%%%%%%%%%%%%%%%%%%%%%%%%%%%

\section{Introduction}
\label{sec:intro}

The cosmic baryon fraction is extremely well determined from
observations of the CMB plus other cosmological probes, with the
latest results from WMAP finding $f_{\rm bar}\equiv \Omega_{\rm
  b}/\Omega_{\rm m}=0.167\pm 0.004$ (Komatsu \etal 2011). However, on
galaxy scales a significant fraction of the baryons are ``missing''.
Stars and cold gas in galaxies account for just $\simeq 8\%$ of the
cosmic baryons (e.g., Bell \etal 2003; Fukugita \& Peebles 2004; Read
\& Trentham 2005), while hot intracluster gas accounts for just
$\simeq 4\%$ (Fukugita \& Peebles 2004).

The vast majority of the cosmic baryons ($\simeq 88\%$) are thought to
be in the form of hot gas in the haloes of galaxies or between
galaxies in the so called warm-hot-intergalactic medium (WHIM) at
temperatures between $10^5$ and $10^7$ K (Cen \& Ostriker
1999). However, only a fraction of these baryons have been detected
(e.g., Bregman 2007, Shull \etal 2011), and the amount of baryons that
reside in hot haloes around the Milky Way and other nearby galaxies is
a subject of current debate (e.g., Grcevich \& Putman 2009; Anderson
\& Bregman 2010).

This raises the question: {\it Why are most of the cosmic baryons in
  hot haloes or the WHIM?}  There are two basic answers: 1) Most of
the baryons accreted into galaxies and were then expelled (into haloes
or the WHIM) by feedback processes (stellar and/or AGN); or 2) Most
baryons never accreted into galaxies in the first place.

The answer to this question is of interest beyond the realm of baryon
accounting. Outflows have been invoked to explain a number of apparent
problems with galaxy formation in a \LCDM context. These include the
predicted central density cusps, which are not observed, but can be
softened with galactic outflows (e.g., Read \& Gilmore 2005;
Mashchenko \etal 2006; Governato \etal 2010; Macci\`o \etal 2012), and
the excess of low-angular momentum material which needs to be removed
in order to produce bulgeless disk galaxies with exponential density
profiles (e.g., Maller \& Dekel 2002; Dutton 2009; Governato \etal
2010; Brook \etal 2011). Thus if galactic outflows are not the
explanation of why galaxy formation is so inefficient, other
mechanisms will need to be found to reconcile \LCDM with observations
on galaxy scales.

A clue to the origin of the missing baryons comes from the fact that
the galaxy formation efficiency\footnote{We define the galaxy
  formation efficiency as the fraction of the cosmically available
  baryons, $\fbar$, that end up as stars and cold gas in a galaxy:
  $\epsilon_{\rm GF}=\Mgal/(\fbar\Mvir)$. } is not a constant. It is
observed to peak at $\epsilon_{\rm GF}\sim 30\%$ in haloes of mass
$\Mvir \sim 10^{12}\Msun$, and declines to both higher and lower
masses (e.g., Hoekstra \etal 2005; Conroy \& Wechsler 2009; Dutton
\etal 2010; Moster \etal 2010; More \etal 2011). In high mass haloes
$\Mvir \gta 10^{12}\Msun$, cooling is progressively more inefficient
(e.g., Blumenthal \etal 1984), which results in the correct
qualitative trend. However, additional heating mechanisms (such as AGN
feedback) are needed in order to reproduce the rapid decline in galaxy
formation efficiency with increasing halo mass (e.g., Croton \etal
2006).  In haloes of mass $10^{10} \lta \Mvir \lta 10^{12}\Msun$ most
of the cosmically available baryons should accrete onto central galaxy
(e.g., Blumenthal \etal 1984; Keres \etal 2009).  In this mass range,
galactic outflows driven by supernovae or stellar feedback are the
leading mechanism for explaining why the galaxy formation efficiencies
are so low, and decline with decreasing halo mass.

Galactic outflows appear ubiquitous in galaxies that are undergoing,
or have recently undergone, significant star formation (e.g., Shapley
\etal 2003; Martin 2005; Tremonti \etal 2007; Weiner \etal 2009; Rubin
\etal 2010; Steidel \etal 2010). At least some of the outflowing gas
is observed to be moving faster than the escape velocity of the
halo. However, measuring outflow mass rates is challenging, and at
present it is not clear how much mass is actually removed (e.g., Rubin
\etal 2010).  Thus, while galactic outflows undoubtedly exist, their
role in determining the baryonic masses of galaxies is unclear.

The scaling relations between baryonic mass, outer rotation velocity
and gas fraction have been used as arguments against galactic outflows
being the explanation for the observed low galaxy formation
efficiencies.
Anderson \& Bregman (2010) argue that outflows should result in a
negative correlation between galaxy mass and stellar mass at fixed
velocity, while no such correlation is observed. They cite this as
strong evidence against galatic outflows.
McGaugh (2012) shows that the efficiency of outflows needs to be
higher in galaxies with higher gas fractions and lower past average
star formation rates. McGaugh (2012) argues that in the context of
feedback models this is apparently puzzling because of the notion that
galaxies with more star formation should have more energy to drive an
outflow. Hence the galaxies that are most efficient at removing
baryons are expected to have the highest star formation efficiencies
and lowest gas fractions. As we show below the resolution of this
puzzle is the fact that it is not just the amount of star formation
that determines the efficiency of feedback. The depth of the potential
well is also important --- it is much easier to remove baryons from
lower mass haloes. In addition McGaugh (2011, 2012) has argued that
the scatter in the BTF is consistent with being zero, which is hard to
explain in a $\LCDM$ context.

It should be noted that until the details of star formation and
feedback are understood it will not be possible to talk of definitive
predictions for the BTF in the \LCDM paradigm. The question we can ask
at the present time is whether the properties of the BTF can be
reproduced in a \LCDM context using {\it plausible} models for star
formation and feedback.  In this paper we address this question using
the semi-analytic disk galaxy formation model of Dutton \& van den
Bosch (2009).  This paper is organized as follows: In \S2 we give a
brief outline of the galaxy formation model; In \S3 we discuss the
correlations between ejection efficiency, galaxy velocity and gas
fraction; In \S4 we discuss the scatter in the baryonic Tully-Fisher
relation; A summary is given in \S5.

%% FIGURE 1
\begin{figure*}
\centerline{
\psfig{figure=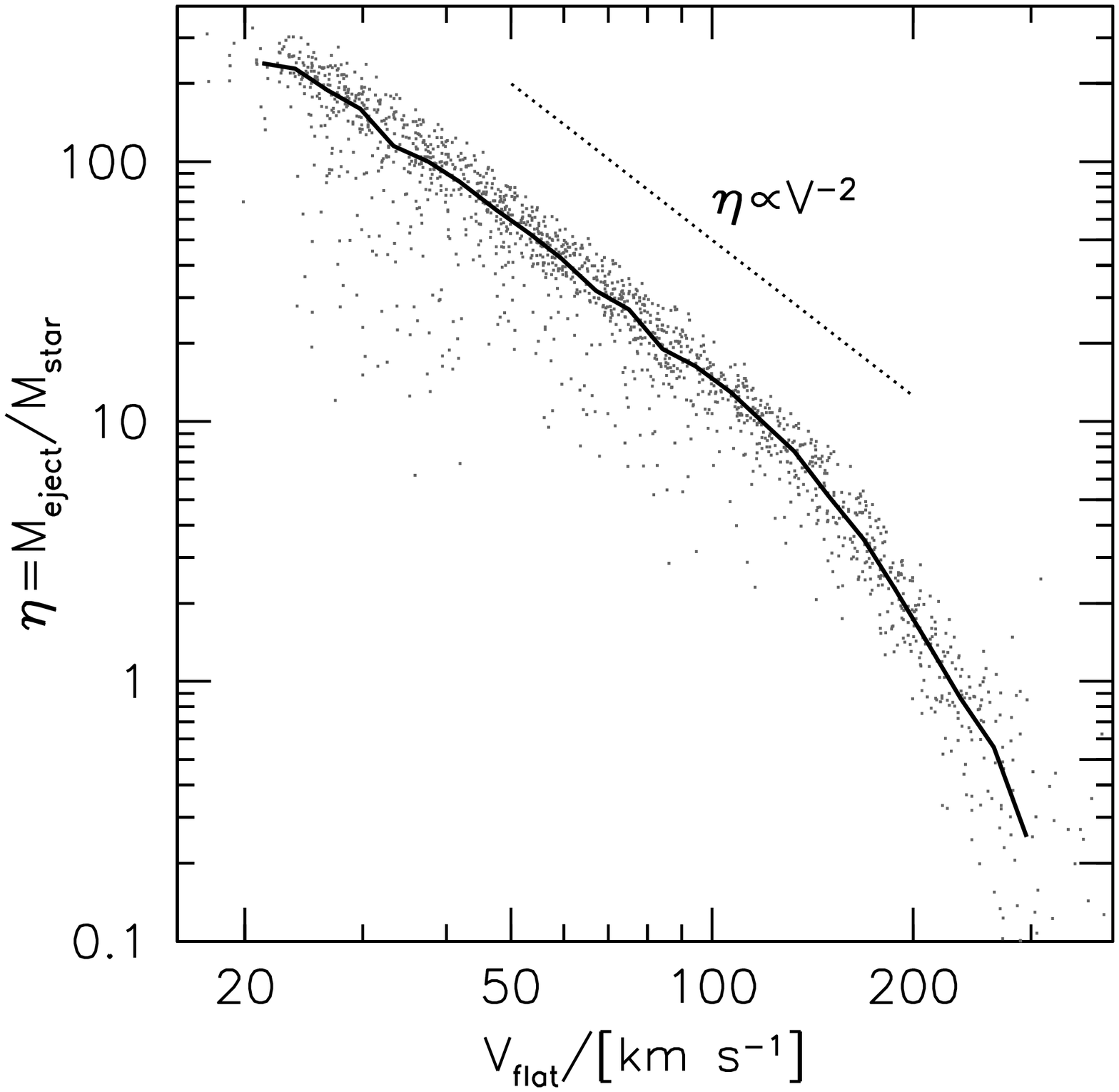,width=0.32\textwidth}
\psfig{figure=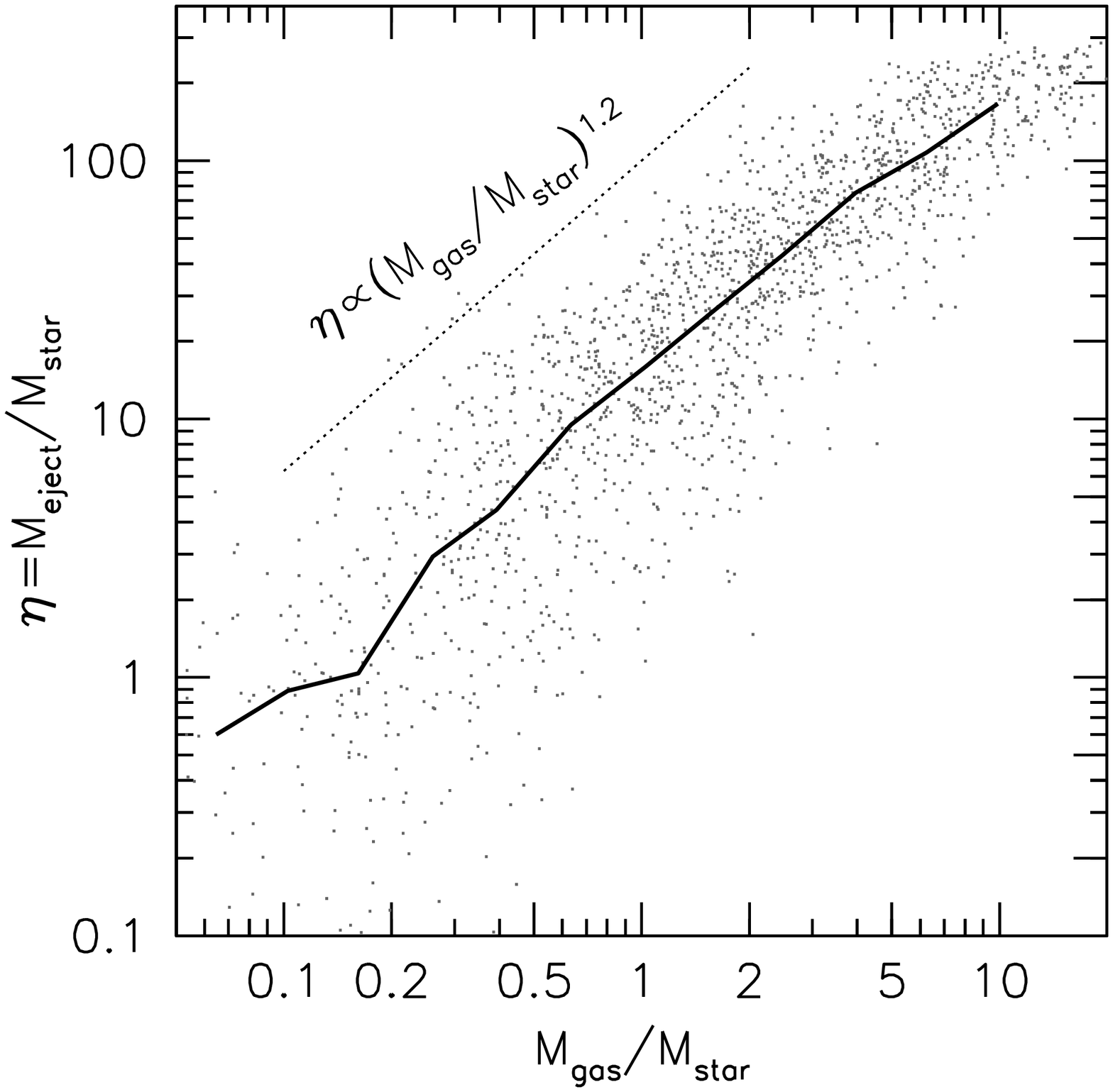,width=0.32\textwidth}
\psfig{figure=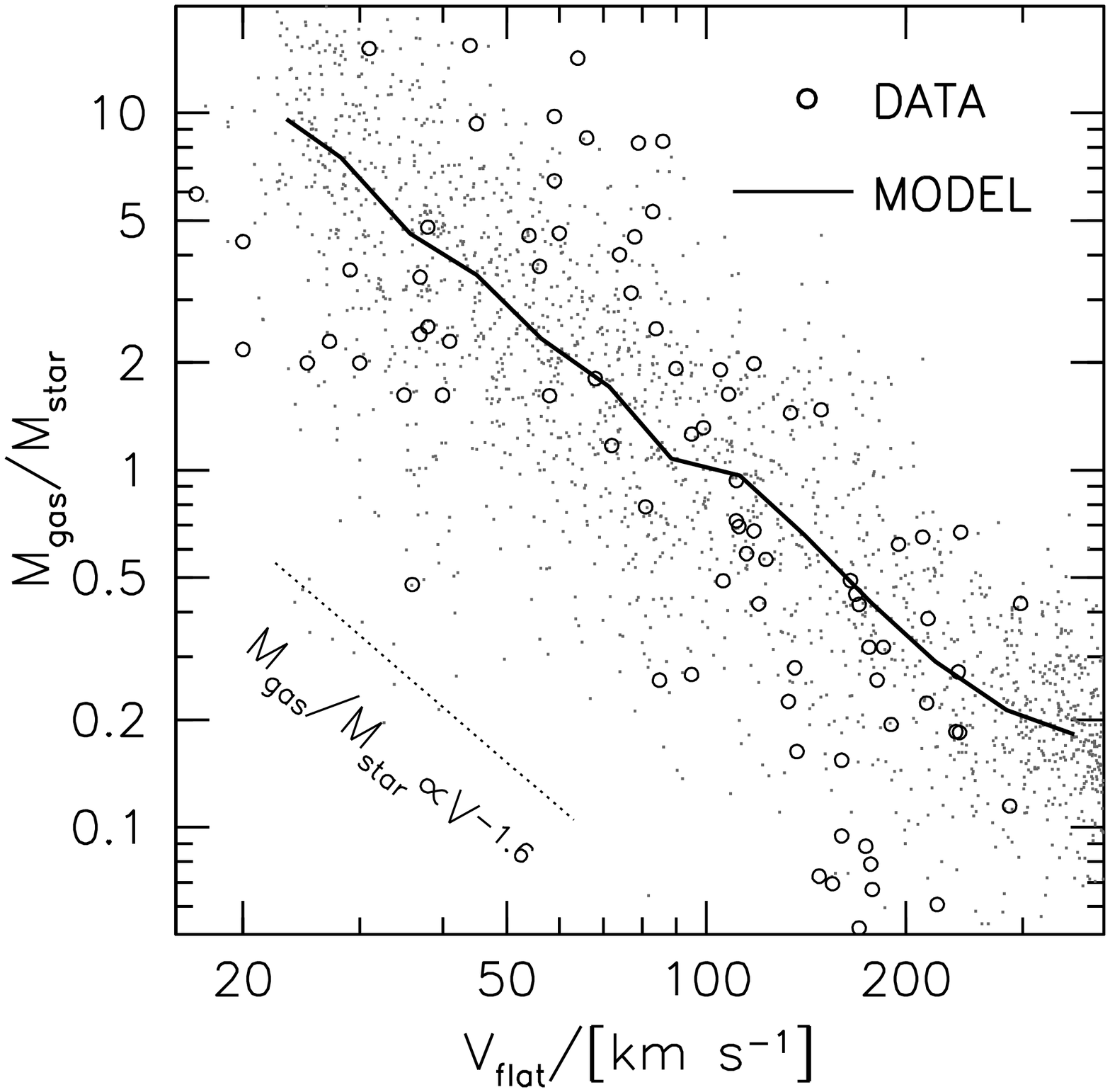,width=0.32\textwidth}}
\caption{Relations between mass loading factor, outer galaxy circular
  velocity and gas-to-stellar mass ratio for a \LCDM based galaxy
  formation model with galactic outflows.  {\it Left:} Feedback is
  more efficient at ejecting baryons from lower mass galaxies, despite
  their lower star formation efficiencies.  {\it Middle:} Feedback is
  more efficient at ejecting baryons from galaxies with higher
  gas-to-stellar mass ratios. {\it Right:} Galaxies with lower
  velocities have higher gas fractions, both in models (points, solid
  line) and observations from Stark \etal (2009) and McGaugh (2012)
  (open points).  This latter correlation explains why middle and left
  panels are mirror images of each other.}
\label{fig:eff}
\end{figure*}

%\vspace{-0.5cm}
\section{Galaxy Formation Model}

In this paper we use a sample of galaxies generated from a \LCDM based
semi-analytic galaxy formation model (Dutton \& van den Bosch
2009). Very briefly, this model follows the evolution of resolved
disks of gas and stars inside smoothly growing dark matter haloes.
The mass accretion histories, dark halo structure and angular momentum
are consistent with cosmological N-body simulations (Bullock \etal
2001a,b; Wechsler \etal 2002; Sharma \& Steinmetz 2005; Macci\`o \etal
2007, 2008). We create a Monte-Carlo sample of 2000 galaxies by
uniformly sampling halo masses from $\log(\Mvir/h^{-1}\Msun) = 9.5$ to
$13.5$, and adding log-normal scatter in halo concentrations
($\sigma_{\ln c}=0.25$), spin parameters ($\sigma_{\ln\lambda}=0.50$),
and angular momentum profile shapes ($\sigma_{\ln \alpha}=0.25$). The
feedback efficiency ($\epsilon_{\rm FB}=0.5$) and angular momentum
losses are tuned to match the galaxy formation efficiency vs halo mass
and galaxy specific angular momentum vs halo mass relations (see
Dutton \& van den Bosch 2012).

The main limitations of this model are discussed in Dutton \& van den
Bosch (2009). Here we highlight two effects that are most likely to
impact the BTF -- the assumptions of smooth mass accretion and that
outflow gas does not return to the galaxy. While smooth accretion is
expected to dominates the build up of spiral galaxies in \LCDM
cosmologies, deviations from this in the form of minor and major
mergers are an unavoidable feature. In our outflow model we only
consider winds that escape the halo, and do not return. It is likely
that some of the ejected gas will return to the halo at a later time
(e.g., Oppenheimer \etal 2010). In addition, it is possible for gas to
escape the disk, but not the halo, and also be reaccreted at a later
time -- which is known as a galactic fountain (e.g., Brook \etal
2012a). All of these effects create scatter in the mass accretion
histories of baryons and dark matter onto galaxies, which one would
nominally expect to result in more scatter in the BTF.

%% FIGURE 2
\begin{figure*}
\centerline{
\psfig{figure=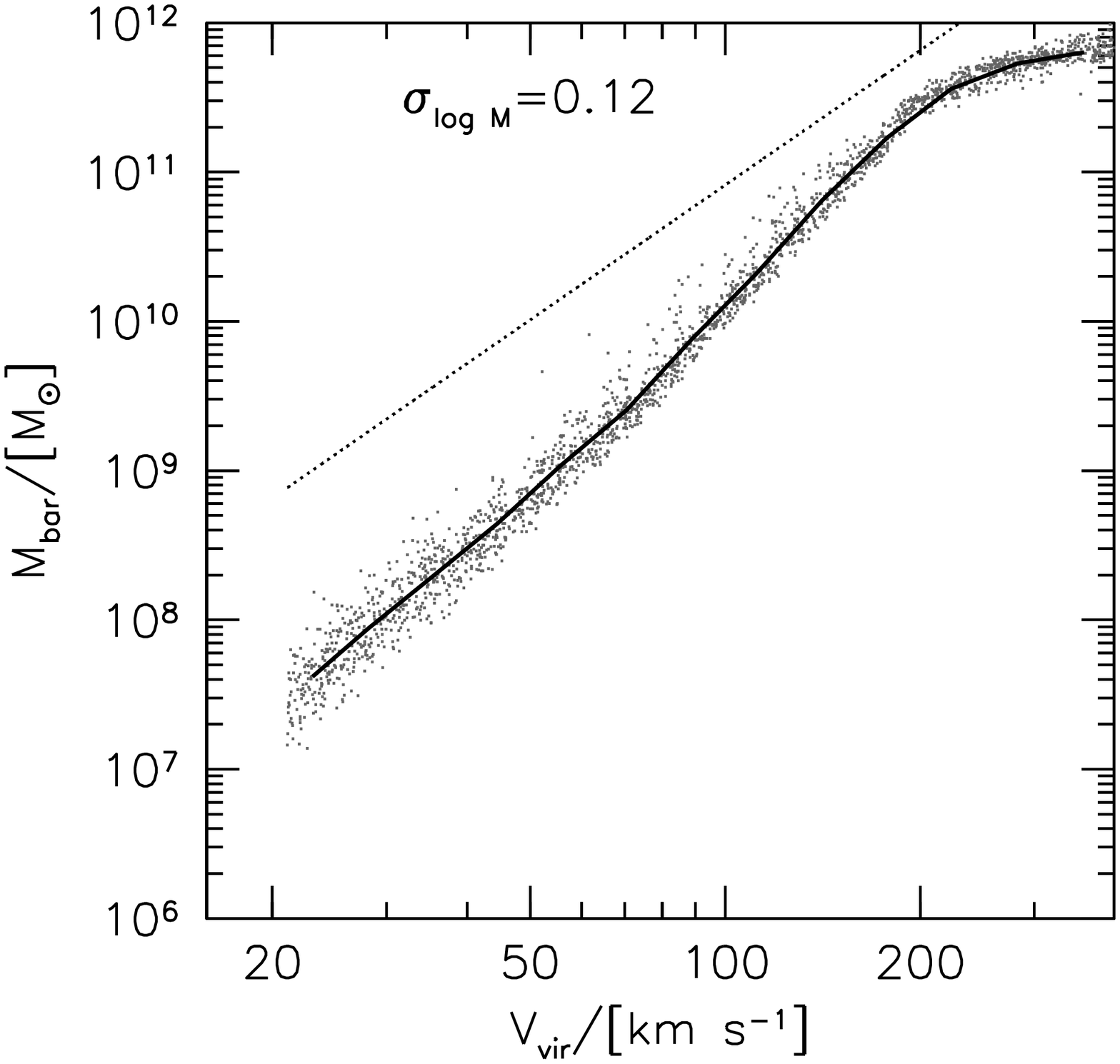,width=0.32\textwidth}
\psfig{figure=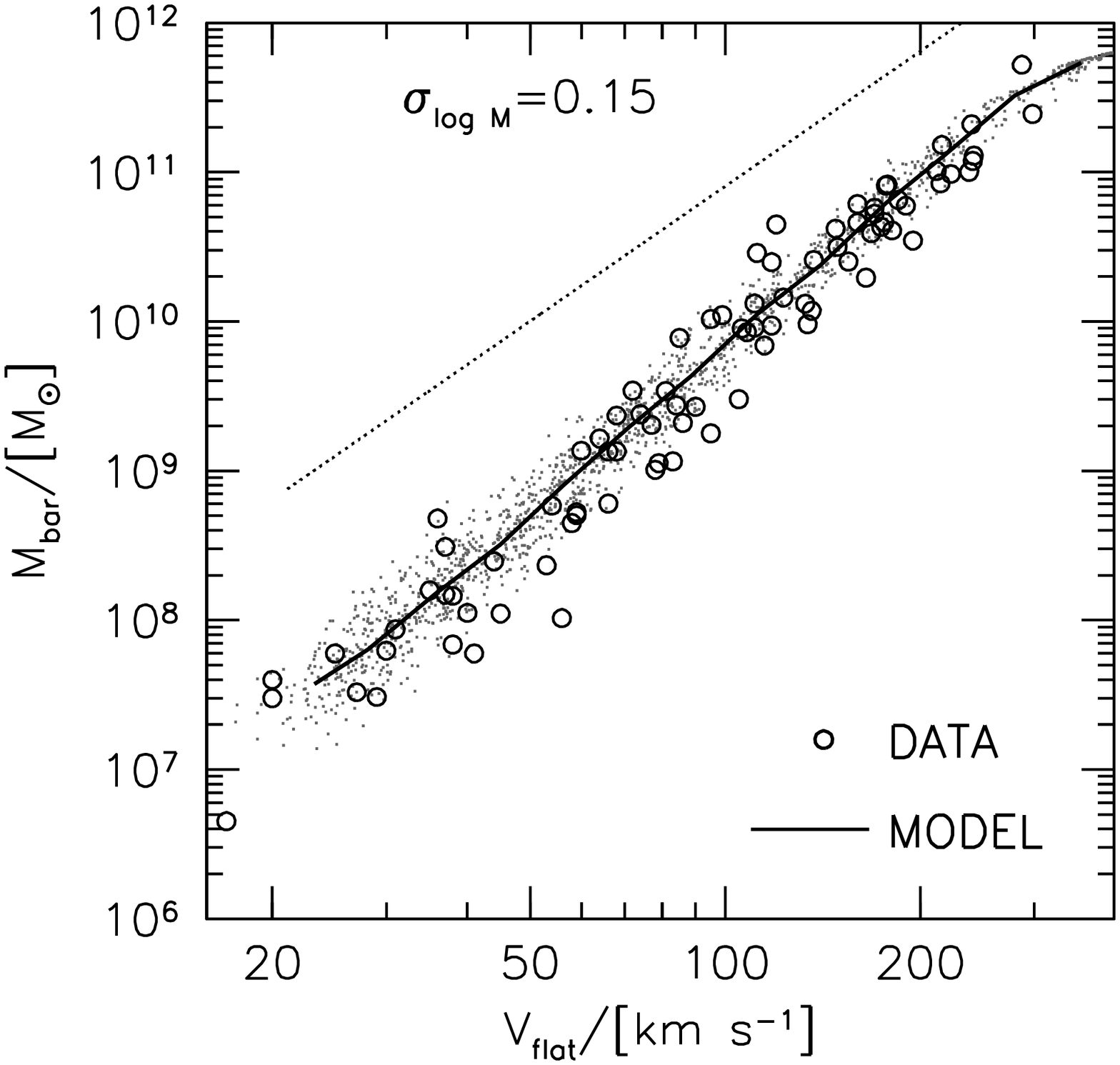,width=0.32\textwidth}
\psfig{figure=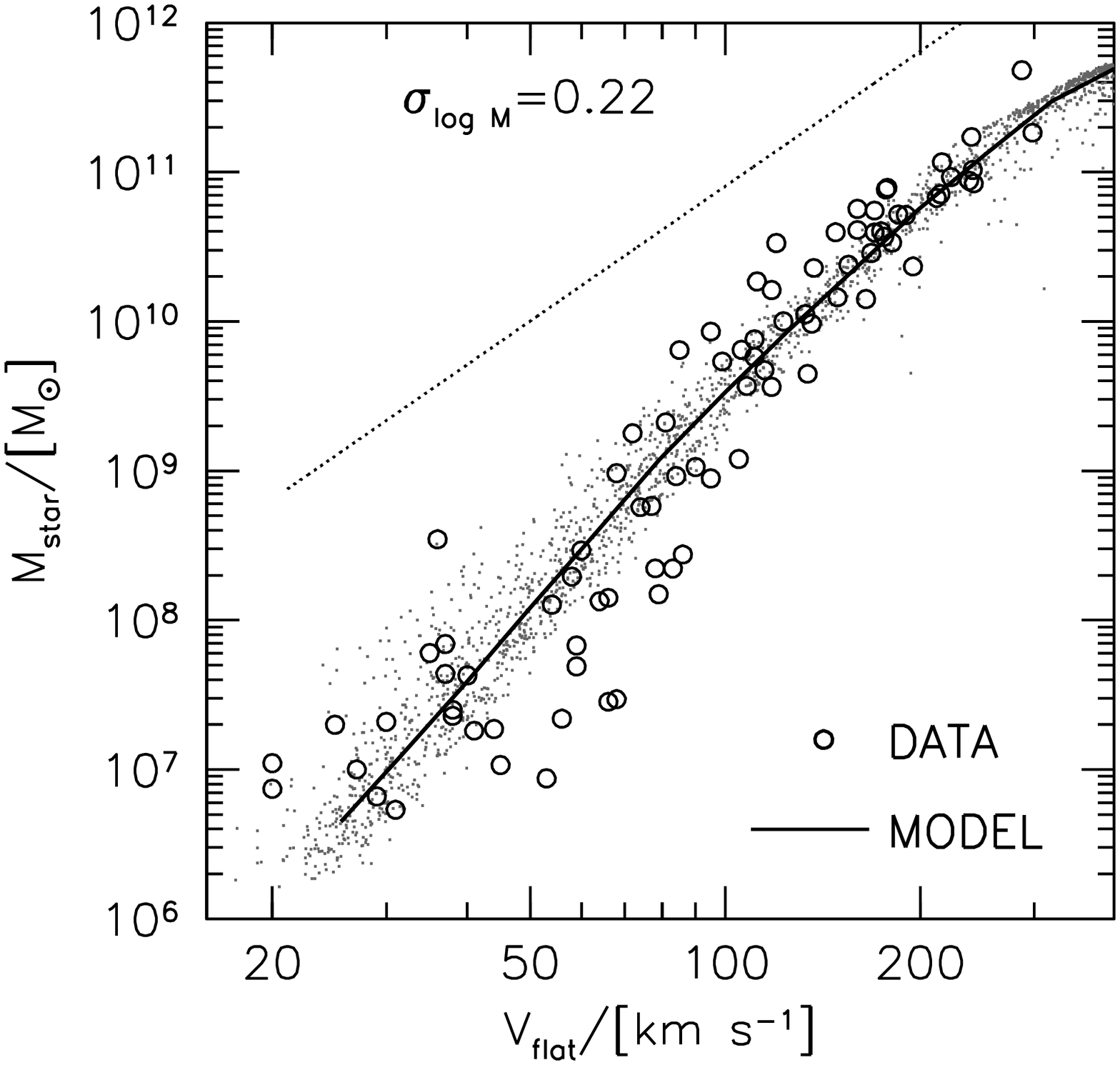,width=0.32\textwidth}}
\caption{Relations between baryonic ($\Mgal$) and stellar ($\Mstar$)
  mass with virial ($\Vvir$) and outer galaxy ($\Vflat$) velocity for
  a \LCDM based galaxy formation model (grey points, solid line) and
  observations from Stark \etal (2009) and McGaugh (2012). The slopes
  of all three relations deviate from the $\fbar\Mvir \propto \Vvir^3$
  scaling for \LCDM haloes, which is indicated by the dotted lines. In
  the models the scatter increases from 0.12 dex for the $\Mgal$ vs
  $\Vvir$ relation to 0.15 dex for $\Mgal$ vs $\Vflat$ (BTF) and 0.22
  dex for $\Mstar$ vs $\Vflat$ (STF).}
\label{fig:VM}
\end{figure*}

%\vspace{-0.5cm}
\section{Ejection efficiency}
\label{sec:fbar}

We define the ejection fraction as the fraction of the cosmically available
baryons that have been ejected from the galaxy:
$ \epsilon_{\rm eject} \equiv M_{\rm eject}/(\fbar\Mvir).$
Similarly we define the stellar mass fraction as the fraction of the
cosmically available baryons that end up in stars (or stellar
remnants):
$\epsilon_{\rm star} \equiv M_{\rm star}/(\fbar\Mvir).$
Note that due to return of gas from stars into the ISM, $\epsilon_{\rm
  star}$ is less than the integral of the star formation rate for any
given galaxy. The ratio of these two quantities is the mass loading
factor: $\eta=\epsilon_{\rm eject}/\epsilon_{\rm star}$, which is a
way to parametrize how ``efficient'' feedback is in ejecting gas
from a galaxy with a given amount of star formation.

The left panel of Fig.~\ref{fig:eff} shows the mass loading factor vs
outer galaxy circular velocity, $\Vflat$\footnote{In our model we
  measure $\Vflat$ at the radius at a radius which encloses 80\% of
  the cold gas.}, for our model. Galaxies with lower circular
velocities have high mass loading factors (i.e., lower mass galaxies
are more efficient at ejecting their baryons). As in the observations
(McGaugh 2012) the effect is not subtle: galaxies with $\Vflat\sim
40\kms$ have $\eta\sim 100$, while galaxies with $\Vflat\sim 200\kms$
have $\eta\sim 1$. The scaling between mass loading factor and
circular velocity $\eta\propto V^{-2}$ (dashed line) results from our
assumptions of energy conservation and that the outflow moves at the
local escape velocity.  We note that momentum driven winds (e.g.,
Murray \etal 2005) are expected to result in $\eta\propto V^{-1}$,
while constant wind velocity models (e.g., Springel \& Hernquist 2003)
result in a constant mass loading factor.  Thus as long as the wind
velocity scales with the galaxy velocity, the lower energy/momentum
input from stars/supernovae in lower mass galaxies is more than
compensated by the shallower potential wells.

The middle panel of Fig.~\ref{fig:eff} shows the mass loading factor
vs gas-to-stellar mass ratio, $\Mgas/\Mstar$. Where $\Mgas$ is the
mass in cold gas (atomic and molecular). This shows that galaxies with
higher gas fractions have higher mass loading factors, and are thus
more ``efficient'' at ejecting their baryons, in qualitative agreement
with observations (McGaugh 2012). A comparison between the left and
middle panels shows that there is less scatter in the relation between
mass loading factor and galaxy velocity, than between mass loading
factor and gas fraction. This suggests the relation between mass
loading and galaxy velocity (left panel) is more fundamental than the
relation between mass loading and gas fraction (middle panel). The
reason galaxies with higher gas fractions have higher mass loading is
simply a result of the anti-correlation between gas fraction and
galaxy velocity (right panel):
$(\Mgas/\Mstar)\propto\Vflat^{-1.6}$. As before, the solid line shows
the median of the model (grey points) which is in good agreement with
the observations (open circles) from Stark \etal (2009) and McGaugh
(2012).

This raises the question:{\it Why do lower mass galaxies have, on
  average, higher gas fractions?} Observationally we know that lower
mass galaxies are on average less dense (e.g., Kauffmann \etal 2003),
and that lower density galaxies are less efficient at turning gas into
stars (Kennicutt 1998). This leads naturally to higher gas fractions
in lower mass galaxies. However, to reproduce the observations in
detail requires, in addition to the standard Schmidt-Kennicutt star
formation law, a threshold density for star formation (van den Bosch
2000).
On the theory side, a correlation between galaxy density and galaxy
mass occurs naturally in a \LCDM context. The simplest disk galaxy
formation model (in which the galaxy formation efficiency and spin
parameters are constant) results in disk sizes $R_{\rm d} \propto
\Mgal^{1/3}$ and thus disk densities $\Sigma_{\rm d} \propto
\Mgal^{1/3}$ (e.g., Mo, Mao, \& White 1998). Including outflows
typically results in shallower size-mass relations (e.g., Dutton \&
van den Bosch 2009), and thus an even stronger mass - gas density
relation.

%\vspace{-0.5cm}
\section{The baryonic Tully-Fisher Relation}
\label{sec:btf}

The baryonic Tully-Fisher relation (BTF) is the relation between the
baryonic mass of a galaxy, $\Mgal$ (stars and cold gas), and the
rotation velocity at large galactic radii, typically referred to as
$\Vflat$. It is an extension of the original Tully-Fisher relation
which is a correlation between \hi line-width and galaxy luminosity
(Tully \& Fisher 1977).  The BTF was first studied by McGaugh \etal
(2000), and subsequently by numerous authors, both observationally
(e.g., Bell \& de Jong 2001; McGaugh 2005; Geha \etal 2006;
Avila-Reese \etal 2008; Begum \etal 2008; Stark \etal 2009;
Trachternach \etal 2009; Gurovich \etal 2010; Hall \etal 2011;
Catinella \etal 2012); and theoretically (e.g., Dutton \& van den
Bosch 2009; de Rossi \etal 2010; Dutton \etal 2011; Piontek \&
Steinmetz 2011; Trujillo-Gomez \etal 2011; Brook \etal 2012b). In this
section we discuss aspects of the slope and scatter in the context of
$\LCDM$.

It has been argued that the ``true'' BTF should include the
contribution of ionized gas (Gnedin 2011). However, this is both hard
to measure and hard to model. In this paper we restrict the baryonic
mass (of both our models and the data) to be that of the stars and
cold gas in a galaxy. Since the vast majority of the ionized gas will
be at radii beyond the \hi disk, the baryonic mass that we use,
$\Mgal$, is close to the baryonic mass within the \hi radius of the
galaxy. As such, $\Mgal$, is expected to be more strongly correlated
to $\Vflat$ than the total baryonic mass inside the virial radius of
the dark matter halo, $\Mbar$. Indeed, in our model the relation
between $\Mbar$ and $\Vflat$ has a slightly larger scatter than the
regular BTF. Likewise the relation between $\Mbar$ and $\Vvir$ has
smaller scatter than the relation between $\Mbar$ and $\Vflat$.

%\vspace{-0.5cm}
\subsection{Slope and zero point of the BTF in LCDM}
In \LCDM the underlying origin of the BTF is the $\Mvir \propto
\Vvir^3$ relation of dark matter haloes. This relation has no scatter
by definition. Predicting the BTF from this relation requires
understanding how baryonic mass is related to virial mass
($f_g=\Mgal/\Mvir$); and how galaxy velocity is related to virial
velocity: $f_V=\Vflat/\Vvir$. Galaxy formation efficiencies depend on
the details of gas cooling, feedback and recycling -- none of which
can be predicted from first principles. The relation between galaxy
and halo velocities is better constrained (thanks to cosmological
N-body simulations), but it too depends on a couple of unknown
factors: the response of the halo to galaxy formation, and the galaxy
formation efficiency. Because the enclosed dark matter fractions
increase with increasing galactio-centric distance, these two unknowns
are minimized by using velocities measured at large radii.  Given
there are no unique predictions for the BTF in \LCDM cosmologies, and
are unlikely to be so for some time, its main utilization is likely to
be as a constraint to galaxy formation models, and in particular
models for feedback. {\it Indeed the feedback efficiency and angular
  momentum loss in our model have been tuned to match the galaxy
  formation efficiency as a function of halo mass.}

%% FIGURE 3
\begin{figure*}
\centerline{
\psfig{figure=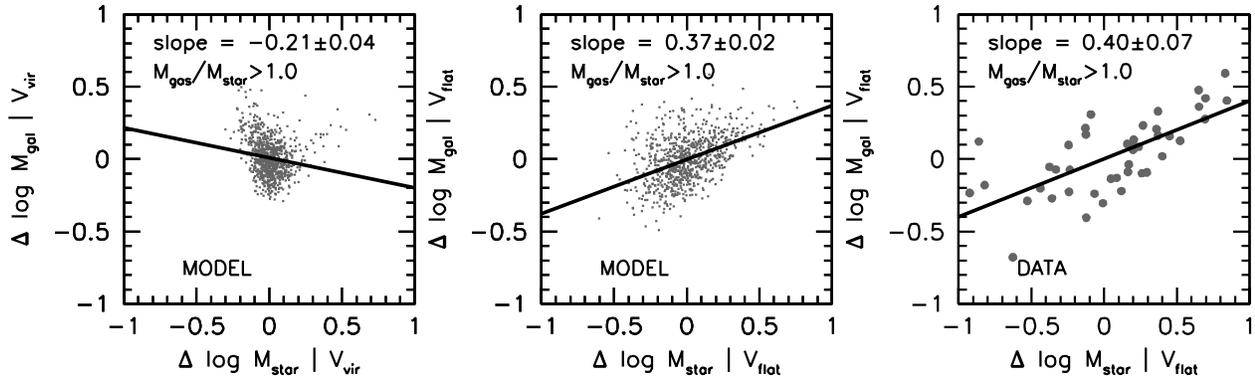,width=0.95\textwidth}}
\caption{Correlations between residuals of the baryonic TF and stellar
  mass TF relations for gas rich galaxies ($\Mgas/\Mstar > 1$).  At
  fixed dark matter halo virial velocity, $\Vvir$, lower stellar
  masses result in higher baryonic masses (left panel), as might be
  expected by the lower energy available to drive an outflow. However,
  when computed at fixed galaxy rotation velocity, $\Vflat$, this
  correlation has the opposite sign (middle panel), with a slope in
  agreement with observations (right panel).}
\label{fig:dVM_gasrich}
\end{figure*}

%\vspace{-0.5cm}
\subsection{Scatter in the BTF}
The intrinsic scatter in the BTF is observed to be small and is
consistent with zero. This is a potential problem for \LCDM (McGaugh
2011, 2012).  Fig.~\ref{fig:VM} shows three Tully-Fisher relations
from our galaxy formation model: Baryonic mass vs virial velocity
(left); Baryonic mass vs galaxy velocity (BTF, middle); and Stellar
mass vs galaxy velocity (STF, right).  All relations have small, but
non-negligible, scatter of $\sim 0.1-0.2$ dex, with the BTF having a
scatter of 0.15 dex. There is also a velocity dependence to the
scatter in all three relations. For the BTF the scatter ranges from
$\simeq 0.20$ dex at $\Vflat\sim 30\kms$ to $\simeq 0.11$ dex for
$\Vflat \sim 200 \kms$.

{\it What is the source of the scatter?} In our models the majority
($\simeq 73\%$) of the scatter in the BTF comes from variation in the
concentration of the dark matter halo, which mostly effects
$\Vflat/\Vvir$. The scatter in halo concentrations is constrained by
cosmological simulations (e.g., Macci\`o \etal 2008), and so there is
little freedom to change this in the context of $\LCDM$.  The
remainder of the scatter comes from variation in the halo spin
parameter, which in our model effects the galaxy formation
efficiency. We note that while variation in the halo spin parameter is
the primary source of variation in the distribution of baryons in
galaxies (i.e., galaxy sizes), this does not significantly effect
$\Vflat$ because we are measuring circular velocity at a radius which
encloses most of the baryons.  Since there are likely other sources of
scatter that are not taken into account in our models, we expect that
our models provide a lower limit to the BTF scatter in $\LCDM$.

{\it What is the intrinsic scatter of the observed BTF?}  The observed
scatter in the BTF is $\simeq 0.24$ dex (McGaugh \etal 2011; Hall
\etal 2011), and is dominated by measurement errors on baryonic masses
(McGaugh 2011; Foreman \& Scott 2011).  Unfortunately, the errors have
uncertainties, so we currently do not have a robust measurement of the
intrinsic scatter. Nevertheless, McGaugh (2012) finds the intrinsic
scatter to be $< 0.15$ dex, which our model is {\it just} consistent
with.

%% FIGURE 4
\begin{figure*}
\centerline{
\psfig{figure=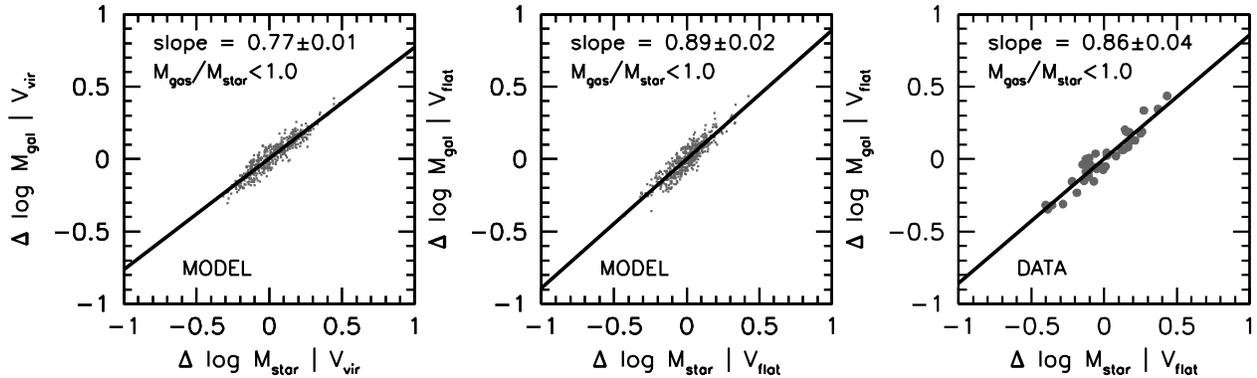,width=0.95\textwidth}}
\caption{As Fig.~\ref{fig:dVM_gasrich} but for gas poor galaxies
  ($\Mgas/\Mstar < 1$). There is a strong correlation between
  residuals of the baryonic TF and stellar TF relations, in both
  models and data. This is a trivial consequence of the fact that
  $\Mstar \simeq \Mgal$ in these galaxies.}
\label{fig:dVM_gaspoor}
\end{figure*}

Thus the intrinsic scatter of the BTF has the potential to be a
powerful constraint on galaxy formation models.  In order to make
progress a BTF sample needs to be constructed for which the
measurement errors are controlled to be smaller than the scatter one
is trying to measure.
There are two primary sources of measurement errors: distance
uncertainties and stellar mass uncertainties. Rotation velocity errors
are typically small when resolved \hi rotation curves are obtained,
and gas masses can be measured reliably.  Distance uncertainties can
be minimized by using galaxies at large enough distances such that
peculiar velocities are not important (i.e., $\gta 100$ Mpc). But
obtaining resolved \hi rotation curves for such galaxies is currently
a challenge, and may have to wait until the next generation radio
telescopes. \hi line widths or H$\alpha$ rotation curves can be
measured for large ($\sim 1000$'s) samples of galaxies (e.g., Courteau
\etal 2007; Hall \etal 2012), but these are not as straightforward to
interpret as $\Vflat$.  For gas poor spiral galaxies, stellar mass
uncertainties are likely to remain the largest source of error, but it
is possible for them to be accurate to $\sim 0.05$ dex (Gallazzi \&
Bell 2009), assuming one knows the form of the stellar initial mass
function.  For gas rich galaxies, the contribution of molecular gas is
the largest source of error, but these galaxies are thought to be
dominated by atomic gas which can be reliably measured.  Thus, in
principle, it should be possible to control measurement uncertainties
to less than 0.1 dex, and thus determine if the intrinsic scatter of
the BTF is smaller than the 0.15 dex of our model.

%\vspace{-0.5cm}
\subsection{Correlation between BTF and STF}
Anderson \& Bregman (2010) used the scatter in the BTF to test the
idea that galactic outflows are responsible for the low observed
galaxy formation efficiencies.
The idea is that for a galaxy in a given dark matter halo, a galaxy
that forms more stars will have more energy/momentum available to
drive an outflow, and thus should remove a higher fraction of its
baryons. Thus at fixed velocity, there should be a negative
correlation between the baryonic mass and stellar mass.  Anderson \&
Bregman (2010) found no such correlation, and thus argued this was
strong evidence against galactic outflows being responsible for the
low galaxy formation efficiencies.

We investigate the validity of this reasoning using BTF and STF
relations from a semi-analytic model. Fig.~\ref{fig:dVM_gasrich} shows
the correlations between the mass residuals of these BTF and STF
relations.  We show results using TF relations constructed using both
virial velocity of the dark matter halo, $\Vvir$, (which is not
observable for individual galaxies), and the velocity in the outer
part of a galaxy, $\Vflat$, (which is observable for individual
galaxies). We split the models into gas rich and gas poor, since gas
poor galaxies are (trivially) expected to have a positive correlation
between baryonic mass and stellar mass (Fig.~\ref{fig:dVM_gaspoor}).
The upper left panel shows that at fixed $\Vvir$, gas rich galaxies do
indeed have a negative correlation between baryonic mass and stellar
mass. However, at fixed $\Vflat$, the correlation has the opposite
sign (middle panel), with a slope in good agreement with observations
from Stark \etal (2009) and McGaugh \etal (2012) (right panel). This
change in slope is a result of scatter in the relation between
$\Vflat$ and $\Vvir$. Thus the simple reasoning used by Anderson \&
Bregman (2010) to argue against outflow models, while correct in
principle, is not valid in practice.

%%%%%%%%%%%%%%%%%%%%%%%%%%%%%%%%%%%%%%%%%%%%%%%%%%%%%%%%%%%%%%%%%%%%%%
%% SECTION X: SUMMARY
%%%%%%%%%%%%%%%%%%%%%%%%%%%%%%%%%%%%%%%%%%%%%%%%%%%%%%%%%%%%%%%%%%%%%%
%\vspace{-0.5cm}
\section{Summary}
\label{sec:conc}

We have used a \LCDM based galaxy formation model to investigate the
observable signatures of galactic outflows in the baryonic
Tully-Fisher relation (BTF). We summarize our results as follows:

\begin{itemize}

\item Observations indicate that galaxies with lower star formation
  efficiencies and higher gas fractions have higher ejection
  efficiencies (e.g., McGaugh 2011).  We show that these trends can be
  explained by energy driven feedback models.

\item In our model feedback is more efficient in galaxies with lower
  circular velocities (and shallower potential wells). This is in spite
  of the significantly lower star formation efficiencies in lower
  velocity galaxies.

\item In our model, as well as observations, lower velocity galaxies
  have higher gas fractions. This results in a (non-causal)
  correlation between ejection efficiency and gas fraction, such that
  ejection efficiencies are higher in galaxies with higher gas fractions.

\item Lower mass galaxies are predicted to have higher gas fractions.
  This is a generic prediction for galaxy formation in $\LCDM$, which
  is strengthened by feedback.

\item The scatter our model BTF is $\simeq 0.15$ dex, and is mostly
  due to variations in the dark matter concentration parameter. While
  this scatter is significantly smaller than the observed scatter of
  $\simeq 0.24$ dex, most of the observed scatter is due to
  measurement uncertainties. McGaugh (2012) finds the intrinsic scatter
  in the BTF is $< 0.15$ dex, which our model is only {\it just} consistent
  with.

\item In principle, future observations of the BTF could be made where
  the measurement uncertainties are controlled to less than 0.1 dex,
  and thus to provide stringent constraints to the intrinsic scatter,
  and \LCDM based galaxy formation models.

\item In our model, at fixed virial velocity ($\Vvir$), gas rich
  galaxies with lower stellar masses have higher baryonic masses. This
  is consistent with the idea that less star formation should result
  in less energy (or momentum) available to drive an outflow (e.g.,
  Anderson \& Bregman 2010). However, at fixed galaxy velocity
  ($\Vflat$), the correlation is of the opposite sign due to the
  scatter in $\Vflat/\Vvir$. Furthermore the slope of the correlation
  in our model is in agreement with observations.
\end{itemize}

In summary, we find that there is currently no conflict between the
observed baryonic Tully-Fisher relation (slope, scatter, and residual
correlations) and predictions of \LCDM based models in which galaxy
formation efficiencies are determined by galactic outflows.

\vspace{-0.5cm}
\section*{Acknowledgements} 
We thank Stacy McGaugh, St\'ephane Courteau, Andrea Macci\`o and Frank
van den Bosch for valuable discussions.

%%%%%%%%%%%%%%%%%%%%%%%%%%%%%%%%%%%%%%%%%%%%%%%%%%%%%%%%%%%%%%%%%%%%%%
%%  REFERENCES
%%%%%%%%%%%%%%%%%%%%%%%%%%%%%%%%%%%%%%%%%%%%%%%%%%%%%%%%%%%%%%%%%%%%%% 

\label{lastpage}

\end{document}